          [1999/12/02 1.01 (PWD)]
\documentclass[letterpaper,twocolumn]{esapub} 

\usepackage{natbib}
\usepackage{psfig}

\title{Coalescence of a strange star with a black hole}
\author{William H. Lee}
\affil{Instituto de Astronom\'{\i}a, Universidad Nacional Aut\'{o}noma 
de M\'{e}xico}
\author{Jon Nix}
\affil{University of Wisconsin--Madison}
\author{W\l odzimierz Klu\'{z}niak}
\affil{Copernicus Astronomical Centre}

\begin{document}

\keywords{}

\maketitle

\begin{abstract}
We present the first numerical results on the binary coalescence of a
quark star with a black hole, obtained with a 3-D Newtonian smooth
particle hydro (SPH) code. The star is initially represented by 17,000
particles modeling a self--gravitating fluid with the equation of
state $P= (\rho - \rho_0)c^2/3$, and the black hole by a point mass
with an absorbing boundary at the Schwarzschild radius.  As in similar
calculations carried out for a stiff polytrope, the stellar core
survives the initial episode of mass transfer, but here an accretion
disk is clearly formed as well.
\end{abstract}

\section{Introduction}

Binary coalescence of compact objects must give rise to gravitational
radiation emission which is expected to be observed with such
instruments as LIGO or VIRGO (e.g. Thorne~1995). Neutron stars
coalescing with black holes may also be the source of r--process
nuclei and of gamma--ray bursts (Lattimer and
Schramm~1976). Coalescence of any compact objects (black holes,
neutron stars, quark stars) may release enough energy to power a
fireball giving rise to some cosmological gamma--ray bursts
(Paczy\'nski~1986), and the coalescence of strange (quark) stars may
be especially promising (Haensel et al.~1991), provided that no quark
matter is spilled into the environment with dire consequences for the
birth rate of neutron stars (Caldwell and Friedman~1991,
Klu\'zniak~1994).

More detailed motivation and further references can be found in Lee
and Klu\'zniak~(1999a), where the SPH coalescence of a tidally locked
neutron star and a black hole is discussed.  Lee and
Klu\'zniak~(1999b) discuss the coalescence of a tidally locked stiff
polytrope with a black hole, while SPH simulations of non--rotating
neutron stars in binary coalescence with black holes have been
presented by Lee~(2000).

\section{Numerical method and initial conditions} \label{method}
We have used the Smooth Particle Hydrodynamics (SPH) method
(Monaghan~1992) for our calculations. Our Newtonian code is three
dimensional and we include a term in the equations of motion that
simulates the loss of angular momentum to gravitational radiation in
the binary (this is calculated in the quadrupole approximation for a
point--mass binary). We also compute the gravitational radiation
waveforms emitted during the coalesence, in the quadrupole
approximation. The strange star is modeled with $N=17000$ SPH
particles, and has mass $M_{SS}=1.4 M_{\odot}$ and radius
$R_{SS}=11$~km. The black hole is modeled as a point mass producing a
Newtonian potential, and we model accretion onto it by placing an
absorbing boundary at the Schwarzschild radius
$r_{Sch}=2GM_{BH}/c^{2}$. For the strange star we take the equation of
state to be $P=c^{2}(\rho-\rho_{0})/3$, with $\rho_{0}=5 \times
10^{17}$~kg~m$^{-3}$.

To perform a dynamical coalescence calculation, we place the binary
components in a close Keplerian orbit at a small separation (a few
stellar radii) and give them the corresponding orbital velocities for
point masses. The strange star is initially spherical and has no spin
as seen from an external, inertial reference frame. The orbit decays
through the emission of gravitational waves and because of tidal
instabilites of purely Newtonian origin (Lai, Rasio \& Shapiro 1993).

Here we present the result of one calculation, with an initial mass
ratio $q=M_{SS}/M_{BH}=0.3$ and initial separation
$r_{i}=3R_{SS}$. The dynamical evolution of the system is followed for
approximately 7~ms. 

\section{Results} \label{results}

At the start of the simulation a tidal bulge appears on the strange
star as a result of the presence of the black hole companion. The
orbital decay causes the star to overflow its Roche lobe, beginning
mass transfer within the first orbit. The star is tidally stretched
into a tube of practically uniform thickness, and some matter that
moves towards the black hole winds around it and produces a torus (see
Figure~1). The outer parts of the star move away from the hole as the
stream connecting the two becomes thinner. Small condensations form at
regular intervals along the narrow stream (see panel (d) in Figure~1),
reminiscent of the results of a black hole--neutron star coalescence
for a stiff equation of state (Lee~2000).

\begin{figure}
\centering
\psfig{width=7.5cm,file=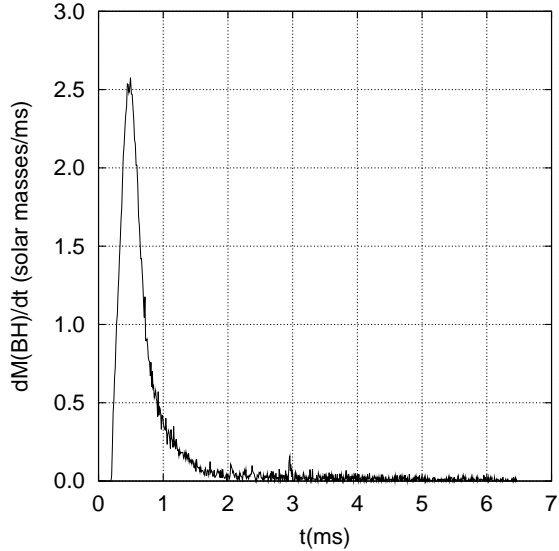,angle=-90,clip=}
\caption{Mass accretion rate onto the black hole as a function of 
time. \label{mdot}}
\end{figure}

\begin{center}
\begin{figure*}
\begin{center}
\psfig{width=14cm,file=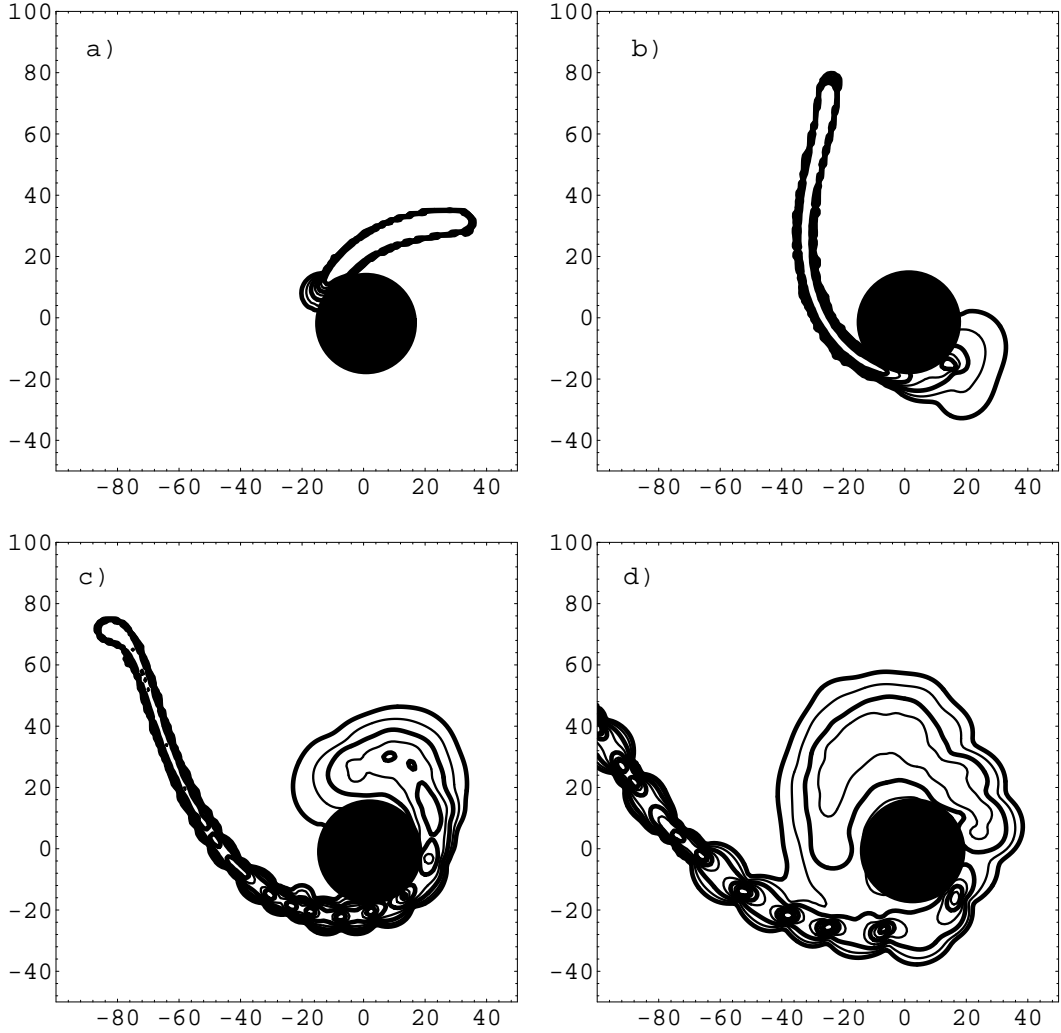,angle=0,clip=}
\caption{Density contour plots in the orbital plane at (a) t=0.85~ms; 
(b) t=1.27~ms; (c) t=1.7~ms; (d) t=2.12~ms. The axes are labeled in
km. All contours are logarithmic and equally spaced every 0.5
dex. Bold contours are plotted at $\log (\rho/\rho_{*})=-4,-3,-2,-1$
(with $\rho_{*}=M_{SS}/R_{SS}^{3}=2.08\times 10^{18}$kg~m$^{-3}$).}
\label{rhoxy}
\end{center}
\end{figure*}
\end{center}

The mass accretion rate onto the black hole shows a large initial peak
at the initial mass transfer episode (see Figure~2), and a subsequent
decay, due to the presence of the accretion torus. The small spikes in
the curve (such as the ones visible at $t\simeq 2.1, 2.4, 3$~ms)
indicate the accretion of small knots along the matter stream coming
from the star.

In the accretion disk around the black hole, the density has dropped
below $\rho_{0}$ and so this matter behaves as pressure--free dust in
orbit about the hole (for reference, in Figure~1
$\log(\rho_{0}/\rho_{*})=-0.62$). However, in the surviving core, flung
to a large distance during the tidal encounter, and in the small
condensations seen in the matter stream connecting the two binary
components, the density remains above this value, and thus the matter
continues to feel hydrodynamic effects. By $t\simeq 5$~ms, there are
0.15~solar masses in the system excluding the black hole, and this is
roughly evenly divided between matter with densities above and below
$\rho_{0}$.

\section{Discussion and conclusions} \label{discussion}

At the end of the calculation we have attempted to obtain an estimate
for the amount matter that could be dynamically ejected from the
system, by computing the total mechanical energy (kinetic +
gravitational potential) of the fluid. We find that $5\times 10^{-3}
M_{\odot}$ could leave the system, and would originate from the
outermost section of the largest blob formed at the end of the tidal
tail. The rest of the matter ($0.15~M_{\odot}$) is gravitationally
bound to the black hole, located practically at the center of mass of
the system, and which has acquired a kick velocity from the asymmetric
interaction ($v_{kick}\simeq 10^{3}$km~s$^{-1}$).

Some of the matter in the accretion torus has enough angular momentum
so as not to be accreted immediatly by the black hole (i.e. it can
remain in orbit in a stable fashion), amounting to
$0.08M_{\odot}$. This fluid would presumably be accreted onto the
black hole on the viscous time scale as angular momentum is
redistributed in the disk.

\section*{Acknowledgments}

This work supported in part by KBN grant 2P03D 004 18, CONACYT
(28987E)and DGAPA--UNAM (IN119998)


\end{document}